\documentclass[aps,pre,nofootinbib,showpacs,twocolumn,10pt]{revtex4}
\usepackage{amssymb,latexsym,mathrsfs}
\usepackage{amsfonts}
\usepackage{amsmath}
\usepackage{graphicx}

\newcommand{\beq}{\begin{equation}}
\newcommand{\eeq}{\end{equation}}
\newcommand{\beqn}{\begin{eqnarray}}
\newcommand{\eeqn}{\end{eqnarray}}
\newcommand{\bearr}{\begin{array}}
\newcommand{\enarr}{\end{array}}

\def\bea{\begin{eqnarray}}
\def\eea{\end{eqnarray}}
\def\ba{\begin{array}}
\def\ea{\end{array}}
\def\n{\nonumber}

\begin{document}
\title{Discontinuous Absorbing State Transition in $(1+1)$ Dimension}
\author{Urna Basu, Mahashweta Basu  and P. K.  Mohanty  }
\email[E-mail address: ]{pk.mohanty@saha.ac.in} 
\affiliation { Theoretical Condensed Matter Physics Division, Saha Institute of Nuclear Physics,
1/AF Bidhan Nagar, Kolkata, 700064 India.}  

\begin{abstract}
 A $(1+1)$ dimensional model of directed percolation is introduced where sites on a tilted square lattice are connected 
to their  neighbours by $N$ channels, operated at both ends by valves  which are either open or closed. The spreading 
fluid is assumed to propagate from any site to the neighbours in a specified direction only through those channels 
which have open valves at both sites. We show that the system undergoes a discontinuous absorbing state transition 
in the large $N$ limit when the number of open valves at each site $n$ crosses a threshold value $n_c=\sqrt N.$ 
Remarkable dynamical properties of discontinuous transitions, like hysteresis and existence of 
two well separated fluctuating phases near the critical point are also observed. The transition is found to be discontinuous in all $(d+1)$ dimensions.
\end{abstract}  

\maketitle
Models of directed percolation (DP)\cite{DPbook} mimics  filtering of fluids through  porous 
media along a given direction.  These models  show a  continuous phase transition from a  
non-percolating (absorbing) to a percolating (active)  state  when  
connectivity of  pores is increased. Numerous physical phenomena, like forest fire\cite{ff},
 epidemics\cite{epi}, transport in random media\cite{rm}, synchronization\cite{sync} and  chromatography
can be understood using  these models. The corresponding critical behaviour 
forms a  universality class of absorbing state phase  transitions,  formally known as the 
DP-class, which  has been  realized \cite{DPexp}convincingly in (2+1) dimension. 

In $(1+1)$ dimension,  the  directed bond percolation\cite{bond} is modelled  as 
a spreading process on a tilted square lattice  where  bonds connect any two 
neighbouring sites with   probability $p.$ A bond represents  an open  
channel connecting two pores. On this random structure, the  spreading agent (or the fluid) 
percolates  along a given direction  through  the bonds (open channels).  The system is 
said to be percolating if the  agent  spreads across  an  infinite cluster of connected  
bonds with finite probability. It is well known that  percolation occurs here as a  
continuous (non-equilibrium) absorbing state phase transition when the  microscopic connection probability
$p$ crosses a threshold value $p_c=0.6447$\cite{Jensen}.  In this  article, we show that, {\it instead},  
if sites are connected by multiple channels, the percolation can occur {\it discontinuously}  
by tuning relevant  parameters.

There has  been a long debate on the  possibility of \textit{discontinuous}  absorbing state 
transition in one dimensional systems. 
It is conjectured by  Hinrichsen\cite{Hinrichsen} that   absorbing  state transition 
can not occur  discontinuously  in one dimensional systems under  generic circumstances 
as strong fluctuations  usually  destabilize the ordered phase. This hypothesis challenged the existence of
the  discontinuous transition in one dimensional triplet creation model  observed by
Dickman and Tom\'e\cite{Tome}, and further supported by Cardoso and 
Fontanari\cite{Fontanari}.  Later studies  show that the  this transition
is actually continuous\cite{Park,Dickman}. Again, the apparent discontinuous  transition 
in  annihilation/fission process\cite{af}, and the spreading process on a diffusing 
background\cite{Wij}, turned out\cite{Hinrichsen} to be  transient phenomena, 
crossing over to a continuous transition after very long time.
Genuine  $1^{st}$ order absorbing state  transitions  are either seen  
in systems with special symmetries \cite{symmetry}  or 
in higher dimensions\cite{BBC,highD}.
 
\begin{figure}[t]
\vspace*{-1cm}
 \includegraphics[width=9cm]{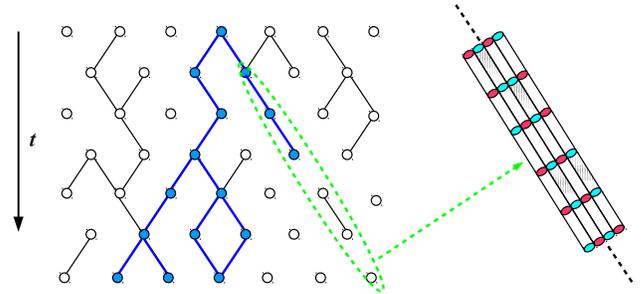}
\vspace*{-2cm}
\caption{(Color online)
Illustration of bond directed percolation in $(1+1)$ dimension with $N=4$ channels.
Each site on a tilted square lattice has $N=4$ valves(discs), of which $n=2$ are open (blue) and the rest are
closed(red). Existence of at least one channel which has open valves at both ends (shaded), represents a bond (solid lines)
between the corresponding sites. Activity  can percolate  only through a bond 
along a  preferred direction (time) activating new sites (filled circles). These bonds  are  shown as
thick blue lines. 
}\label{fig:pic}
\end{figure}

In this Letter we introduce a   multichannel model of  bond directed percolation 
in (d+1) dimensions  where  every pair of neighbouring sites has 
$N$ different channels joining them.  Each  channel contains a valve   at every  site that can 
either be  open or closed   independent of other valves. A bond is said to connect 
two sites if there exists at least one channel joining  them,  which has open valves at both  the  ends. 
Like the usual model of bond directed percolation,  the spreading agent  here propagates in  
a specific direction starting from a  random site.  We show that   the system 
undergoes a discontinuous percolation transition  in the large $N$ limit, even in $(1+1)$ dimension, 
when the number of  open valves $n$ at each site crosses a threshold value $\sqrt N$.  
We believe that these results convincingly establish  the existence of    first 
order  absorbing state  transition in $(1+1)$ dimension.

First let us discuss this multichannel model of  directed percolation in $(1+1)$ dimension in details. 
Its extension to higher dimension  is straightforward.  The sites  on a  tilted square lattice are  
connected to their neighbours by $N$ directed channels (Fig. \ref{fig:pic}).  Each  channel contains 
a valve at every lattice site  so that,   at every site, there are $N$ valves in   total. 
Each valve can either be open or closed  independent of other valves. Flow through a channel 
is possible only when the  channel is {\it open}, $i.e.,$
it has open valves at both ends.  In analogy with the  usual bond percolation process, 
a {\it bond}  is said to be present between  two neighbouring sites when  there exists  at least one open channel 
connecting them. The bond is absent if none of the connecting channels are open. At each site, randomly  chosen $n$ valves  out of $N$ 
are opened, which mimics a porous medium with randomly distributed directed {\it bonds} on  (1+1) dimensional 
tilted square lattice. Like  the usual bond directed percolation problem,  percolation  is 
said to occur here if  there exists a  directed path  which spans the system. 
     
 DP is  often regarded as a dynamical spreading process where the fluid percolates 
in a given direction, interpreted as time.  A similar interpretation for this multichannel model 
follows. In (1+1) dimension, the lattice sites  in horizontal  
and vertical directions   are denoted  by  spatial and time  coordinates  $i=1,2\dots L$ and $t$ 
respectively (Fig. \ref{fig:pic}). From  any arbitrary site $(i,t)$ the spreading fluid  
can independently propagate to the nearest neighbouring sites $(i\pm1,t+1)$ only 
if the  target site is connected to  $(i,t)$  by a bond. In other words, the fluid can reach (activate)
$(i\pm1)^{th}$  site at  time $t+1$ from $(i,t)$ if there is at least one open  channel between 
$(i\pm1,t+1)$ and $(i,t).$  If none of the  sites at $t+1$  gets activated, $i.e.$  
if the number of {\it active sites} at time $t+1$ is zero, which happens when  there is no 
bond connecting  $(i,t)$ and $(i\pm 1,t+1)$ for any $i$, 
the spreading agent can not propagate further. Corresponding configuration at $t+1$ is  
called {\it absorbing}. In this dynamical description, the system  is said to be percolating  
when the fluid keeps spreading for arbitrarily long time. 
Note that when the system is percolating,  there must exist one or more directed paths spanning the $(1+1)$  
dimensional lattice. Thus, this dynamic description is completely equivalent with the 
geometric interpretation of the model discussed in the previous paragraph. Henceforth we continue 
with the dynamical description.

A  continuous phase transition  from the absorbing to active  state  occurs  in the usual  models of bond 
directed percolation  when  the connection probability $p$ is increased  beyond a critical threshold $p_c$. 
The steady state density of active sites $\rho_s$ plays the role of the order parameter as it  takes   
non-zero value only in the active phase 
$p>p_c.$ In  the multichannel  model  introduced here  $n$ plays the role of the control parameter as the 
connectivity between  neighbouring sites increases when more valves are open. Clearly  there   can  not be  
any open  channel for $n=0$,   hence  there are no bonds  available for spreading.  Corresponding  
steady state density of active sites  $\rho_s=0$. Again, for  $n> N/2$ every pair of neighbouring sites 
has at least one open channel connecting them. 
Starting from any random site the spreading agent  surely activates all the sites in the steady state,
resulting in $\rho_s=1.$  Our aim  here  is to find  if the system percolates for  any non-zero $n_c<N/2.$

Undirected bond percolation on similar multichannel models \cite{BBKM}  are known to exhibit 
percolation transition  in  one and higher dimensions when  the number of channels $N\to\infty$. The transition 
occurs discontinuously  as the number of  open valves $n$  crosses  $n_c =  \sqrt N.$  In the present work 
we investigate the possibility of an absorbing state transition  by  varying $n$ along with $N$ 
as $n =  N^\gamma$  where  $0<\gamma <1.$

\begin{figure}
\includegraphics[width=8cm]{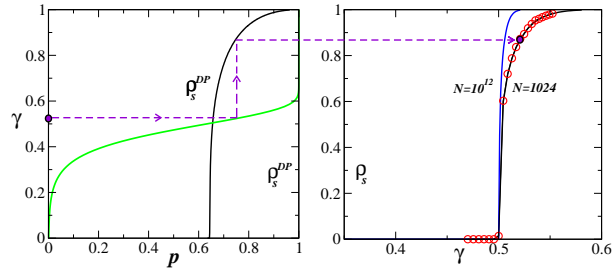}
\caption{(Color online) (a) The steady state   density of active sites $\rho_s^{DP}$ of usual bond 
directed percolation on a  tilted   square 
lattice  with $L=1000$ is plotted against $p$, along with  $\gamma$  versus $p$   [calculated from  Eq. \eqref{eq:pgamma}] 
for  the multi channel model in $(1+1)$ D with $N=1024$. For any value of $\gamma$, which corresponds to a unique $p$ 
(thus  $\rho_s^{DP}$) one can read out $\rho_s$  using \eqref{eq:DP} [following the arrow].  The 
resulting $\rho_s$ for $N=1024$ and  $10^{12}$ are shown as solid lines in (b). Symbols therein correspond 
to $\rho_s$  obtained from  direct  numerical simulation  of  our model  in $(1+1)$ D with $L=1000,$  
 $N=1024$. }\label{fig:joint}
\end{figure}


Since at every lattice site, $n$ valves are 
 chosen randomly from the total $N$ and opened,  which can be done in $C^N_n$ possible 
ways, the probability that there is no common open valve between any two 
neighbouring sites, {\it i.e.}  the probability that neighbouring sites are  {\it not} connected 
by  any open channel, is 
$
q =    C^{N-n}_{n}/ C^N_n .
$
Using Stirling's approximation, $q$ can be expressed in terms of $\nu=n/N$ and $N$, 
\bea
q&=& {(1-\nu)\over \sqrt{1-2\nu}} \exp{[-g(\nu)N]}\label{eq:qN}\\
{\rm where~~}  
g(\nu)&=& -\ln \left[(1-\nu) \left( {1-\nu \over 1-2\nu}\right)^{(1-2\nu)}\right].\n 
\eea 
Note that, here   $p=1-q$  is the  probability that   neighbouring sites  are connected by a bond. 
In  the large $N$ limit, $\nu = N^{\gamma -1}$ vanishes  and   $g(\nu) = \nu^2$ to the leading order in $\nu$.
This results in 
\bea 
p=1-\exp( - N^{2\gamma -1}) .   \label{eq:pgamma}
\eea
Thus,
\bea
\lim_{N\to\infty}  p =  \left\lbrace
 \begin{split} 
&0 &~~~&   {\rm for ~~} \gamma < 1/2  \\
&1-e^{-1} &~~~&  {\rm for ~~}     \gamma = 1/2 \\
&1 &~~~&    {\rm for ~~}   \gamma >1/2.
\end{split}\right.
\label{eq:lim_p}
\eea 
Clearly, the connection probability $p$ shows a discontinuity  at $\gamma_c=\frac 1 2$ in the 
$N\to\infty$  limit  yielding a transition  from  the absorbing  state  (where there are no bonds) to 
an active phase (where all neighbouring sites are connected by bonds).  Thus  for any 
$\epsilon= (\gamma -\gamma_c) >0$ 
the spreading agent   eventually activates  all the sites in steady state resulting in $\rho_s=1.$
This absorbing  state transition  is discontinuous as  the order parameter $\rho_s$ 
changes discontinuously  from $0$ to $1$ as 
$\gamma$  crosses   $\gamma_c=\frac 1 2.$

The order parameter $\rho_s(\gamma;N)$ for  any  finite $N$ can be calculated 
from the knowledge  of the steady state density of active sites  
$\rho_s^{DP}(p)$  of  usual  bond directed percolation  model for a given  connection 
probability $p$, 
\bea 
\rho_s(\gamma;N)=\rho_s^{DP}(p=  1 -  N^{2\gamma-1}).
\label{eq:DP}
\eea
 As  the analytical form of $\rho_s^{DP}(p)$ is not known,  we first  obtain the same  numerically  by simulating 
the usual bond directed percolation problem  on a tilted  square  lattice with $L=1000$ for $0\le p\le 1$. 
$\rho_s(\gamma;N)$ can be read 
out from this data using Eq. \eqref{eq:DP} where  $p=1 -  N^{2\gamma-1}.$  is used.   This procedure is  
illustrated in Fig. \ref{fig:joint}(a) for $N=1024$.   The  resulting $ \rho_s(\gamma;N)$  is shown  
in   Fig. \ref{fig:joint}(b) as a solid line.  The symbols therein  correspond to the  same 
obtained from direct simulation of the multi channel system on a  tilted square lattice 
with $L=1000$ having   $N=1024$ channels.   
It is clear from   this figure that the  transition occurs discontinuously near $\gamma_c=1/2$.  
The order parameter $\rho_s(\gamma;N)$  for $N=10^{12}$, obtained using the 
above procedure, is also shown  in Fig. \ref{fig:joint}(b) to illustrate that, as expected, 
the transition point shifts  towards $\gamma_c=1/2$  as $N\to \infty$.

\begin{figure}
\includegraphics[width=7cm]{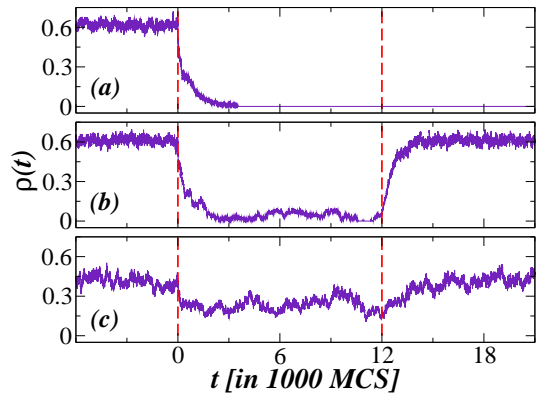}
\caption{
(Color online)$\rho(t)$ near the transition point: The system in steady state with $\gamma = \gamma_c + \epsilon_1$
is quenched to $\gamma= \gamma_c- \epsilon_2$ at $t=0$  and  the $\gamma$  is changed back to its initial 
value after $12000$ MCS. (a) and (b) correspond to our model with $\delta=0, 10^{-5}$  respectively.
(c) shows $\rho(t)$ for conventional DP  near the critical point $p_c$ with  same $\epsilon_1$, $\epsilon_2$
and $\delta.$ Here $\epsilon_1=0.005$ and $\epsilon_2=0.001.$ 
}\label{fig:time}
\end{figure}

The existence of a  discontinuity  in the order parameter $\rho_s$  is clear from the above 
discussions.  However the $1^{st}$ order phase transitions possess several remarkable dynamical 
aspects including hysteresis which distinguish  them from the continuous ones. In the following 
we will show the  numerical evidence of hysteresis,  the most important signature of a 
discontinuous transition. Before going into the details of these studies first we  check   that the 
time-series of the instantaneous density of  active sites  $\rho(t)$
shows  a   sharp drop when the  tuning parameter $\gamma$ is suddenly decreased below 
the critical threshold $\gamma_c$[marked as  $t=0$ in Fig. \ref{fig:time}(a)]. 
Such a finite drop does not occur in  the models of usual directed  
percolation  as  the  order parameter there vanishes continuously at $p_c.$

In any  discontinuous transition,  it is expected that the system  near the critical threshold 
fluctuates  between   the ordered  and the disordered phases leading to interesting  
dynamical behaviour.  However  it is difficult to study these signatures  in a  discontinuous 
absorbing state transition, as once a system  reaches  an absorbing configuration  
it remains there forever. 
To overcome this difficulty, Bidaux et. al. \cite{BBC} have introduced a special technique where  
inactive sites are  spontaneously activated with a 
small rate $\delta.$  This  rate $\delta$ does not change the  discontinuous nature of the 
transition but converts  the absorbing state into a  fluctuating low density state  
allowing one to study  hysteresis and  related dynamical aspects of the 
discontinuous absorbing state transition.  In presence of $\delta$, the transition  in this system 
occurs  from a high density to a low density phase. Figure  \ref{fig:time}(b)  shows the time series 
of $\rho(t)$,  when  the system in steady state at $\gamma=\gamma_c+\epsilon_1=0.505$ ($\epsilon_1=0.005$) is quenched  
to  $\gamma=\gamma_c-\epsilon_2=0.499$  ($\epsilon_2=0.001$) and kept 
there for $12000$ Monte Carlo steps (MCS). In the Monte-Carlo simulation we have used $\delta=10^{-5}.$
Here, $\rho(t)$   drops to the   low density state   soon after the quench ($t=0$)   
and again  jumps back
to the high density phase as $\gamma$ reverts back to the initial value. 
This  behaviour is typical of  $1^{st}$ order phase transitions.  
A similar variation of  the tuning parameter across the critical point of  ordinary directed 
percolation does not  show [Fig. \ref{fig:time}(c)] any visible change in $\rho(t)$. This is 
because of the fact that in   absence of the absorbing state, the  steady state density  does 
not change appreciably across $p_c.$ 

\begin{figure}[tb]
\includegraphics[width=7.5cm]{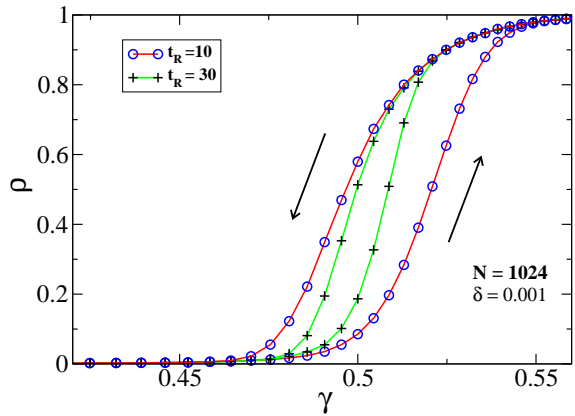}
\caption{(Color online) 
The multichannel model shows hysteresis near the  transition point.
Here activity is created spontaneously with a  small rate $\delta= 10^{-5}$.  In the
Monte-Carlo simulation we have used  $L=1000$, $N=1024.$ The two loops correspond to the relaxation
 times $t_R=10$ (circle) and $t_R=30$ (plus). Arrows indicate the direction in which the loops are generated.  
 }\label{fig:hyst}
\end{figure}

\textit{Hysteresis:} 
To study  hysteresis the system is first  relaxed at $n>\sqrt N$ such that the corresponding 
$\gamma$ is well above the critical  threshold.  Then  $n$  is decreased in steps of $1$  after 
every  $t_R$ Monte-Carlo steps (MCS) until a  sufficiently low value of $\gamma$  below $\gamma_c$ 
is reached. Now  the path is reversed {\it i.e.} $n$  is  increased in steps of $1$, and the hysteresis 
cycle is completed.  In  Fig. \ref{fig:hyst}, we have  shown the  hysteresis curves in the $\gamma$-$\rho$ 
plot for a system of $L=1000, N=1024$ with different $t_R=10,30$.  The fact that  the hysteresis curve 
bounds  a finite  area   reinforces   the claim that this  absorbing state transition is discontinuous. 
As expected this area decreases with  the increase of $t_R$.


 \textit{Site DP:} The model can be  recast  into a different  form  by associating a 
set $S$ of $n$ integers, randomly chosen from  a larger set $\{1,2,\dots N\}$,  to 
each lattice site. In this picture, a bond is said to connect two neighbouring 
sites  $(i,t)$ and $(i\pm1,t+1)$ if $S(i,t) \cap  S(i\pm1,t+1) \ne \O$, where $\O$ is  
the  null set (an empty set without any element).  The   larger set, here, corresponds 
to $N$ valves existing at each site and the smaller set $S$ corresponds  to  the $n$ open valves. In fact, this 
picture is  useful in describing site directed percolation in this multichannel model.  
 
In case of site directed percolation, each site $i$  at time $t$ contains 
a  set $S(i,t)$  which is either a null set $\O$ (inactive)  or a  set  of $n$  integers
randomly chosen  from  $\{1,2,\dots N\}$. A site is said to be active  when  the  
associated set  is non-empty ($S\ne\O $). From any given  configuration at $t$ 
the next configuration at time $t+1$  can be generated as follows.
If both $S(i\pm1,t) =\O$, then $S(i,t+1)$  is assigned the null set $\O$. 
Otherwise, {\it i.e.} when at least one of them  is  not $\O$,   a trial set $T$ of $n$  integers 
is created and  $S(i,t+1)$ is updated as,
\bea
 S(i,t+1)= \left\lbrace 
\begin{split}
T \;\;\; {\rm if} \;\; T\cap  K \ne  \O \\
\O \;\;\; {\rm if}\;\; T \cap  K =  \O  .
\end{split}
\right.
\eea 
where $K$ is one of  $S(i\pm1,t)$ if both are non-empty; otherwise  $K$ is the only non-empty set among  $S(i\pm1,t)$.
Evidently, through this dynamics, new active sites are created only from the existing neighbouring active sites
with probability $p$  which follow   Eq. \eqref{eq:pgamma} when  $n$  varies as $n= N^\gamma$. 
Thus, in  $N\to \infty$ limit, here too one expects  a  discontinuous transition similar to that of the 
bond  percolation model. Details of this study are omitted here.

It is straight forward to  study this multichannel model of directed percolation on any kind of lattice in any $(d+1)$ 
dimension.  In higher dimensions also, one expects  the transition to occur discontinuously 
at $\gamma_c= \frac{1}{2}$ ({\it i.e.} $n_c= \sqrt N$)  as the  connection  probability 
$p$  [Eq. \eqref{eq:pgamma}] itself  becomes discontinuous when  number of channels $N\to \infty$.

In summary,  we have studied  directed percolation in $(d+1)$  dimension on a multi channel system
where   neighbouring sites are connected by $N$ channels, each having an operating  valve at either 
ends. Of these  $N$ valves at each site, $n$ are open. The fluid can percolate in a given direction, 
only through the channels which have  open valves at  both ends. In the  $N\to \infty$ limit, this 
system shows a discontinuous transition from a percolating phase to a non-percolating 
(absorbing) state as   $n$ is decreased below a  threshold $n_c= \sqrt N.$

It is rather surprising that   the absorbing state transition in this  $(1+1)$  dimensional  model 
is  discontinuous. The possibility of such a transition has been rebutted by Hinrichsen \cite{Hinrichsen}, 
who argued that unless there is an unusually robust mechanism  to 
prevent fluctuation from destabilizing the ordered  phase,  it is impossible to  have a discontinuous  
absorbing state transition in generic $(1+1)$ dimensional systems with short range interactions. 
However, this  argument does not seem to apply here, as the discontinuity in the order parameter 
is  the artefact  of the  discontinuity that  appears in the resulting  connection probability  
while varying   the control parameter $\gamma$ continuously across $\gamma_c=1/2$; 
the role of fluctuation is immaterial here.

It is worth mentioning  that  depinning  or synchronization, which are 
described by a multiplicative noise Langevin equation\cite{Munoz} similar to that of DP, 
show discontinuous transitions in certain model systems \cite{syncFPT}. It would be  interesting to 
investigate whether the discontinuous  absorbing state transition  obtained  in the present study  
falls in the realm of their  theory.

  {\it Acknowledgements :}  UB would like to acknowledge thankfully 
the financial support of the Council of Scientific and Industrial Research, India (SPM-07/489(0034)/2007).  
 

\end{document}